\documentclass[a4paper,notab,nolineno, UKenglish,cleveref, autoref, thm-restate]{socg-lipics-v2021}
\usepackage{amsthm}
\usepackage{amsmath}
\usepackage{amssymb}
\usepackage{amsfonts}
\usepackage{bm}
\usepackage{bbm}
\usepackage{enumerate}
\usepackage{graphicx}
\usepackage{hyperref}
\usepackage{import}
\usepackage{multirow}
\usepackage{multicol}
\usepackage{psfrag}
\usepackage[dvipsnames]{xcolor}
\usepackage{nicefrac}

\usepackage{threeparttablex}
\usepackage{tabularx}
\usepackage{edtable}

\DeclareFontFamily{U}{FdSymbolA}{}
\DeclareFontShape{U}{FdSymbolA}{m}{n}{<-> s * [1] FdSymbolA-Book}{}
\DeclareFontShape{U}{FdSymbolA}{m}{b}{<-> s * [1] FdSymbolA-Medium}{}
\DeclareSymbolFont{fdsymbols}{U}{FdSymbolA}{m}{n}
\SetSymbolFont{fdsymbols}{bold}{U}{FdSymbolA}{m}{b}
\DeclareMathSymbol{\bdiamond}{\mathbin}{fdsymbols}{130}

\usepackage{newunicodechar}
\newunicodechar{├}{\resizebox{0.15em}{1.1em}{|}\hspace{-0.1em}-}
\newunicodechar{└}{\raisebox{0.4em}{\resizebox{0.15em}{0.6em}{|}}\hspace{-0.1em}-}

\newcommand{\Oplus}{\ensuremath{\vcenter{\hbox{\scalebox{1.2}{$\oplus$}}}}}

\DeclareMathOperator*{\cproduct}{\operatorname*{\Oplus}}

\newcommand{\eps}{\varepsilon}

\newcommand{\mm}[1]{{\boldsymbol{#1}}}
\newcommand{\bb}[1]{\mathbbm{#1}}
\newcommand{\cc}[1]{\mathcal{#1}}

\newcommand{\bR}{\bb{R}}

\newcommand{\mmp}{\mm{p}}
\newcommand{\mmq}{\mm{q}}

\newcommand{\mmu}{\mm{u}}
\newcommand{\mmv}{\mm{v}}

\newcommand{\mmx}{\mm{x}}
\newcommand{\mmy}{\mm{y}}

\def\bpm{\begin{pmatrix}}
\def\epm{\end{pmatrix}}
\def\bal{\begin{aligned}}
\def\eal{\end{aligned}}

\newcommand{\bi}{\begin{itemize}}
\newcommand{\ei}{\end{itemize}}
\newcommand{\beq}{\begin{equation}}
\newcommand{\eeq}{\end{equation}}

\newcommand{\ben}{\begin{enumerate}}
\newcommand{\een}{\end{enumerate}}

\usepackage{algorithm}
\usepackage{algorithmic}
\usepackage{subcaption}

\title{Smallest Intersecting and Enclosing Balls} %

\author{
Jiaqi Zheng }{
Department of Computer Science, National University of Singapore}{
jiaqi@u.nus.edu}{
https://orcid.org/0009-0005-7493-8274}{}

\author{
Tiow-Seng Tan }{
Department of Computer Science, National University of Singapore}{
tants@comp.nus.edu.sg}{}{}

\authorrunning{J. Zheng and T.S. Tan} %

\Copyright{Jiaqi Zheng and Tiow-Seng Tan} %

\ccsdesc[500]{Theory of computation~Computational geometry}
\ccsdesc[500]{Theory of computation~Approximation algorithms analysis}
\ccsdesc[500]{Theory of computation~Convex optimization}

\keywords{Geometric optimization, smallest intersecting ball, approximation algorithm} %

\makeatletter
\begingroup\endlinechar=-1\relax
       \everyeof{\noexpand}%
       \edef\x{\endgroup\def\noexpand\homepath{%
         \@@input|"kpsewhich --var-value=HOME" }}\x
\makeatother
\def\overleafhome{/tmp}%

\begin{document}

\maketitle

\begin{abstract}
    We study the smallest intersecting and enclosing ball problems in Euclidean spaces for input objects that are compact and convex. They link and unify many problems in computational geometry and machine learning.
    We show that both problems can be modeled as zero-sum games, and propose an approximation algorithm for the former. Specifically, the algorithm produces the first results in high-dimensional spaces for various input objects such as convex polytopes, balls, ellipsoids, etc.
\end{abstract}

\section{Introduction}

\begin{figure}[t]
\centering
\begin{subfigure}[t]{.3\linewidth}
    \centering
    \includegraphics[height=1.5in]{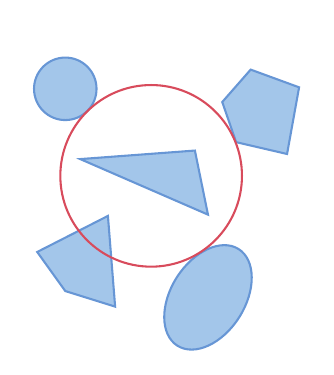}
\end{subfigure}
~
\begin{subfigure}[t]{.3\linewidth}
    \centering
    \includegraphics[height=1.5in]{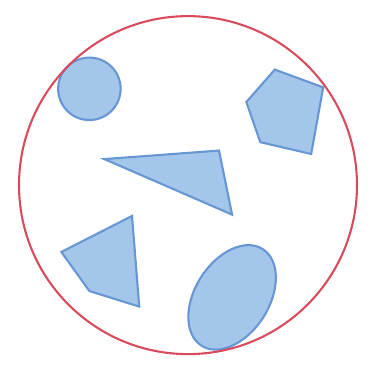}
\end{subfigure}
\caption{Examples of the problems in 2D spaces, where the blue objects are the input and red circles are the solutions. {\em Left:} the smallest intersecting ball. {\em Right:} the smallest enclosing ball. }\label{fig:2d_examples}
\end{figure}

Given $n$ convex compact objects $\Omega_1, \dots, \Omega_n$ in $d$-dimensional Euclidean space, the {\em smallest intersecting ball} (SIB) problem is to find a ball with the smallest radius $r^*$ that {\em intersects} every $\Omega_i$, while the {\em smallest enclosing ball} (SEB) problem is to find the ball with the smallest radius $R^*$ that {\em encloses} every $\Omega_i$.
See Figure~\ref{fig:2d_examples} for 2D examples of these two problems.

\begin{figure}[t]
\centering
\begin{subfigure}[t]{.23\linewidth}
\centering
\includegraphics[height=1.3in]{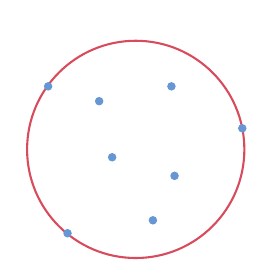}
\end{subfigure}
~
\begin{subfigure}[t]{.3\linewidth}
\centering
\includegraphics[height=1.3in]{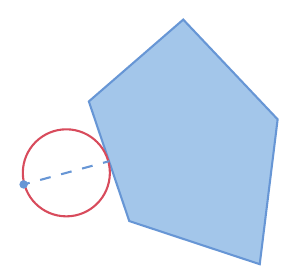}
\end{subfigure}
~
\begin{subfigure}[t]{.4\linewidth}
\centering
\includegraphics[height=1.3in]{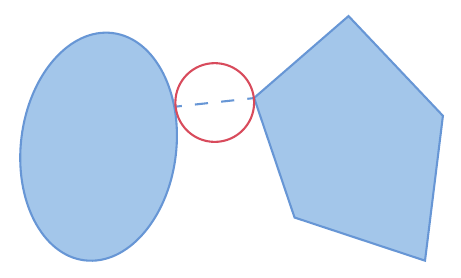}
\end{subfigure}
\caption{Many faces of the SIB problem. {\em Left:} the SEB of a point set. {\em Middle:} the nearest point (Euclidean projection) in a convex set. {\em Right:} the shortest connector (minimum distance).}\label{fig:manyfaces}
\end{figure}

The SEB problem has attracted significant attention in the past decades \cite{sescoreset:badoiu2003smaller,seb:yildirim2008,sublinearopt:CHW2012}, whereas the SIB problem is less discussed and the understanding of SIB lags behind that of SEB.
In earlier research \cite{siblowdim:BJMR1991,sibtheory:MNV2012}, SIB are usually considered a variant of SEB.
Indeed, they are identical when the input are singleton sets.
Nevertheless, as the complexity of the input structure increases, the divergence between these two problems becomes more evident and the SIB problem manifests greater versatility.
This is demonstrable even when there are only two objects, $\Omega_1$ and $\Omega_2$:
when $\Omega_1$ is a compact convex set and $\Omega_2$ is a single point, the SIB problem is equivalent to finding the nearest point (Euclidean projection) of $\Omega_2$ in the region of $\Omega_1$, and $r^*$ is half the distance from $\Omega_1$ to $\Omega_2$;
when $\Omega_1$ and $\Omega_2$ are both convex compact sets, the SIB problem becomes finding the shortest line segment (a.k.a. the shortest connector) that connects these two sets, and $r^*$ is half the minimum distance between them. 
The dual problem of minimum connector is to find the hyperplane that separates $\Omega_1$ and $\Omega_2$ with the largest margin \cite{distance:Achiya2006}, which corresponds to the support vector machine problems in machine learning \cite{svm:BB2000,pd:GJ2009}.
See Figure~\ref{fig:manyfaces} for examples of SIB in different cases.

Given the diversity of the SIB problem, one can reasonably anticipate that it poses more substantial computational challenges than SEB. 
Indeed, numerous algorithms have been proposed for solving the SEB problem, including exact and approximation algorithms \cite{sebexact:Welzl1991,sescoreset:nielsen2009approximating}, using optimization or coreset techniques \cite{sebopt:BS2000,sescoreset:badoiu2003smaller}, and in parallel or streaming settings \cite{sublinearopt:CHW2012,sebstreaming:CP2014}, but for SIB, most algorithms are merely designed for solving it in fixed dimensions \cite{siblowdim:BJMR1991,siblowdim:JMB1996}.

In this work, we endeavor to narrow the gap in the understanding of these two problems.
We show that both the SIB and SEB problems can be modeled as two-player zero-sum games, which is inspired by the seminal work of Clarkson et.~al.~\cite{sublinearopt:CHW2012} in sublinear optimization.
Based on the new formulation, we propose the first approximation algorithm for the SIB problem in arbitrary dimensions in the unit-cost RAM model, which leverages recent advances in symmetric cone problems~\cite{scmwu:CLPV2023, pdscp:ZVTL2024}.
Additional details on the SIB algorithm can be found in the full-length preprint~\cite{sibarxiv}.
Software implementing the algorithm is available at~\cite{libsib}.

\section{SIB and SEB as Zero-Sum Games}

We use $\cproduct$ to denote concatenations of vectors, or Cartesian products of sets and vector spaces. For instance, $\cproduct_{i=1}^n \mmu_i$ denotes the concatenation of $n$ vectors, namely $(\mmu_1, \dots, \mmu_n)$. 
Consider the zero-sum game $\min_{\mmp\in \cc{P}} \max_{\mmq\in \cc{Q}} f(\mmp, \mmq)$. We say $(\mmp^*, \mmq^*)$ is a {\em Nash equilibrium} iff $f(\mmp^*, \mmq^*) \le f(\mmp, \mmq^*), \forall \mmp \in \cc{P}$ and $f(\mmp^*, \mmq^*) \ge f(\mmp^*, \mmq), \forall \mmq \in \cc{Q}$.
Moreover, $f(\mmp^*, \mmq^*)$ is the {\em value} of the game.
Let $\cc{V} := \cproduct_{i=1}^n \Omega_i$, $\cc{X}$ be the convex hull of the input, and $\cc{Y}$ defined~as:
\[
    \cc{Y} := \Big\{\cproduct_{i=1}^n (\mmy_i, s_i) \in \cproduct_{i=1}^n \bR^{d + 1} : \|\mmy_i\| \le s_i, \forall i \in [n] \text{ and } \sum_{i=1}^n s_i = 1 \Big\},
\]
which can be viewed as the Cartesian product of $n$ Euclidean balls whose radii sum to one.

\begin{theorem}
The SIB problem can be modeled as the following zero-sum game:
\[
    \min_{(\mmx, \mmv_1, \dots, \mmv_n)\in \cc{X} \times \cc{V}} \
    \max_{\mmy \in \cc{Y}} \
    \Big( \cproduct_{i=1}^n \begin{pmatrix}
        \mmx - \mmv_i \\
        0
    \end{pmatrix} \Big)^\top \mmy.
\]
A Nash equilibrium (denoted as $(\mmx^*, \mmv_1^*, \dots, \mmv_n^*, \mmy^*)$) of the game always exist, and the value of the game is $r^*$. 
The ball $B(\mmx^*, r^*)$ is intersecting every $\Omega_i$,~and~$\mmv_i^* \in B(\mmx^*, r^*) \cap \Omega_i$.
\end{theorem}

\begin{theorem}
The SEB problem can be modeled as the following zero-sum game:
\[
    \min_{\mmx\in \cc{X}} \
    \max_{(\mmy, \mmv_1, \dots, \mmv_n) \in \cc{Y} \times \cc{V}} \
    \Big( \cproduct_{i=1}^n \begin{pmatrix}
        \mmx - \mmv_i \\
        0
    \end{pmatrix} \Big)^\top \mmy.
\]
A Nash equilibrium (denoted as $(\mmx^*, \mmy^*, \mmv_1^*, \dots, \mmv_n^*)$) of the game always exist, and the value of the game is $R^*$. The ball $B(\mmx^*, R^*)$ is enclosing every~$\Omega_i$.
\end{theorem}

It is worth noting that the SIB game is a bilinear zero-sum game, where the objective function is linear for both min- and max-player. 
On the other hand, the SEB game is not bilinear as the function is not linear (and neither convex nor concave) for the max-player.

\section{Algorithms}

Unlike the SEB problem that is extensively studied in the literature, most algorithms for the SIB problem are designed for fixed dimensions with limited types of input objects such as convex polytopes \cite{siblowdim:JMB1996} and axis-aligned bounding boxes \cite{siblowdim:LK2010}. 
The only result for SIB in high-dimensional space is restricted to input of Euclidean balls that are pairwise disjoint~\cite{sibhighdim:SA2015}. 

Benefit from our new formulation for the SIB problem, we can utilize techniques for bilinear zero-sum games to design an approximation algorithm for general input objects in arbitrary dimensions.
Specifically, we say $(\mmx, r)$ is an $(1 + \eps)$-approximate solution of the SIB problem if the ball $B(\mmx, r)$ intersects every $\Omega_i$ and $r \le (1 + \eps) r^*$.
The algorithm works as follows: in each iteration, we update $\mmy$ using an online optimization algorithm over $\cc{Y}$, and let $(\mmx, \mmv_1, \dots, \mmv_n)$ be the best response in $\cc{X}\times \cc{V}$ against $\mmy$.
Then it can be shown that the average point of the past iterates converges to an approximate Nash equilibrium of the SIB game, which provides an approximate solution of the SIB problem.

\begin{theorem}
Let $D$ be the diameter of the input and let $R = {D \over r^*}$. 
Suppose the best response can be computed in $O(S)$ time.
Then there is an iterative algorithm that computes an $(1 + \eps)$-approximate solution of the SIB problem with running time $O({R^2 (S + nd) \log n \over \eps^2})$.
\end{theorem}

The complexity results in the unit-cost RAM model for the SIB problem with specific input are shown in Table~\ref{table}. 
See \cite{sibarxiv} for detailed analyses of our results.
On the other hand, no existing algorithm can find Nash equilibria for the SEB game due to its non-bilinear nature.
We hope for further advancement on the SEB problem under the new formulation. 

\begin{table}[t]
\caption{Summary of the results for the SIB problem}\label{table}
\setlength\extrarowheight{2pt}
\vspace{-1.2em}
\begin{center}
\begin{tabular}{|l||c|c|}
\hline
\multicolumn{1}{|c||}{Input Objects} & Previous Work & Our Result \\
\hline\hline
Convex Polytopes & \scalebox{.95}{$O(M)^\dag$~\cite{siblowdim:JMB1996}} & \scalebox{.95}{${O}({R^2(N + nd)\log n \over \eps^2})$}\\
Axis-Aligned Bounding Boxes & \scalebox{.95}{$O(n)^\dag$~\cite{siblowdim:LK2010}} & \scalebox{.95}{${O}({R^2 nd \log n \over \eps^2})$} \\
Euclidean Balls & \scalebox{.95}{$O({n \over \eps^{(d-1)/2}})$~\cite{sibhighdim:SA2015}} & \scalebox{.95}{${O}({R^2 nd \log n \over \eps^2})$} \\
Ellipsoids & - & \scalebox{.95}{${O}(nd^\omega + {R^2 nd^2 \log n \over \eps^2})$} \\
\hline
\end{tabular}%
\end{center}
\vspace{-0.3em}
\scalebox{0.85}{
\hspace{-.4em}\parbox{1.17\linewidth}{
\rule{10em}{.5pt}

Note: $d$ is the dimensionality. $n$ is the number of objects. $M$ is the total number of points. $N$ is the number of nonzeros in the input. $R$ is the ratio between $D$ and $r^*$. $\omega$ is the matrix multiplication exponent.

$^\dag$ Running time of exact algorithms for problems in fixed dimensions.
}
}
\end{table}

\ifx\homepath\overleafhome
\bibliography{reference}
\else
\bibliography{reference}

\begin{thebibliography}{10}

\bibitem{svm:BB2000}
Kristin~P. Bennett and Erin~J. Bredensteiner.
\newblock Duality and geometry in {SVM} classifiers.
\newblock In {\em Proceedings of the Seventeenth International Conference on Machine Learning}, pages 57--64, 2000.

\bibitem{siblowdim:BJMR1991}
Binay~K. Bhattacharya, Sreesh Jadhav, Asish Mukhopadhayay, and Jean-Marc Robert.
\newblock Optimal algorithms for some smallest intersection radius problems.
\newblock In {\em Proceedings of the Seventh Annual Symposium on Computational Geometry}, SCG '91, pages 81--88, New York, NY, USA, 1991.

\bibitem{sescoreset:badoiu2003smaller}
Mihai Bădoiu and Kenneth~L. Clarkson.
\newblock Smaller core-sets for balls.
\newblock In {\em SODA}, volume~3, pages 801--802, 2003.

\bibitem{scmwu:CLPV2023}
Ilayda Canyakmaz, Wayne Lin, Georgios Piliouras, and Antonios Varvitsiotis.
\newblock Multiplicative updates for online convex optimization over symmetric cones.
\newblock {\em arXiv preprint arXiv:2307.03136}, 2023.

\bibitem{sebstreaming:CP2014}
Timothy~M Chan and Vinayak Pathak.
\newblock Streaming and dynamic algorithms for minimum enclosing balls in high dimensions.
\newblock {\em Computational Geometry}, 47(2):240--247, 2014.

\bibitem{sublinearopt:CHW2012}
Kenneth~L. Clarkson, Elad Hazan, and David~P. Woodruff.
\newblock Sublinear optimization for machine learning.
\newblock {\em J. ACM}, 59(5), 2012.

\bibitem{distance:Achiya2006}
Achiya Dax.
\newblock The distance between two convex sets.
\newblock {\em Linear Algebra and its Applications}, 416(1):184--213, 2006.

\bibitem{pd:GJ2009}
Bernd G{\"a}rtner and Martin Jaggi.
\newblock Coresets for polytope distance.
\newblock In {\em Proceedings of the twenty-fifth annual symposium on Computational geometry}, pages 33--42, 2009.

\bibitem{sebopt:BS2000}
Bernd G{\"a}rtner and Sven Sch{\"o}nherr.
\newblock An efficient, exact, and generic quadratic programming solver for geometric optimization.
\newblock In {\em Proceedings of the sixteenth annual symposium on Computational geometry}, pages 110--118, 2000.

\bibitem{siblowdim:JMB1996}
Shreesh Jadhav, Asish Mukhopadhyay, and Binay Bhattacharya.
\newblock {An Optimal Algorithm for the Intersection Radius of a Set of Convex Polygons}.
\newblock {\em Journal of Algorithms}, 20(2):244--267, 1996.

\bibitem{siblowdim:LK2010}
Maarten Löffler and Marc {van Kreveld}.
\newblock Largest bounding box, smallest diameter, and related problems on imprecise points.
\newblock {\em Computational Geometry}, 43(4):419--433, 2010.

\bibitem{sibtheory:MNV2012}
Boris Mordukhovich, Nguyen~Mau Nam, and Cristina Villalobos.
\newblock The smallest enclosing ball problem and the smallest intersecting ball problem: Existence and uniqueness of solutions.
\newblock {\em Optimization Letters}, 7(5):839--853, Apr 2012.

\bibitem{sescoreset:nielsen2009approximating}
Frank Nielsen and Richard Nock.
\newblock Approximating smallest enclosing balls with applications to machine learning.
\newblock {\em International Journal of Computational Geometry \& Applications}, 19(05):389--414, 2009.

\bibitem{sibhighdim:SA2015}
Wanbin Son and Peyman Afshani.
\newblock {Streaming Algorithms for Smallest Intersecting Ball of Disjoint Balls}.
\newblock In {\em Theory and Applications of Models of Computation}, pages 189--199, 2015.

\bibitem{sebexact:Welzl1991}
Emo Welzl.
\newblock Smallest enclosing disks (balls and ellipsoids).
\newblock In {\em New Results and New Trends in Computer Science: Graz, Austria, June 20--21, 1991 Proceedings}, pages 359--370. Springer, 2005.

\bibitem{seb:yildirim2008}
E~Alper Yildirim.
\newblock Two algorithms for the minimum enclosing ball problem.
\newblock {\em SIAM Journal on Optimization}, 19(3):1368--1391, 2008.

\bibitem{libsib}
Jiaqi Zheng.
\newblock {LIBSIB: A C++ library for computing smallest intersecting balls in arbitrary dimensions}, 2024.
\newblock URL: \url{https://github.com/orzzzjq/libsib}.

\bibitem{sibarxiv}
Jiaqi Zheng and Tiow-Seng Tan.
\newblock Approximation algorithms for smallest intersecting balls.
\newblock {\em arXiv preprint arXiv:2406.11369}, 2024.

\bibitem{pdscp:ZVTL2024}
Jiaqi Zheng, Antonios Varvitsiotis, Tiow-Seng Tan, and Wayne Lin.
\newblock A primal-dual framework for symmetric cone programming.
\newblock {\em arXiv preprint arXiv:2405.09157}, 2024.

\end{thebibliography}
\fi

\end{document}